\documentclass[
twocolumn, preprintnumbers, amsmath,amssymb]{revtex4}

\usepackage{graphicx}
\usepackage{dcolumn}
\usepackage{bm}

\newcommand\ee{\end{equation}}
\newcommand\be{\begin{equation}}
\newcommand\eea{\end{eqnarray}}
\newcommand\bea{\begin{eqnarray}}
\newcommand{\sfrac}[2]{{\textstyle\frac{#1}{#2}}}
\newcommand\di{\partial}

\begin{document}


\title{An equivalence principle for scalar forces
}

\author{Lam Hui}
\email{lhui@astro.columbia.edu}

\author{Alberto Nicolis}
\email{nicolis@phys.columbia.edu}

\affiliation{%
Physics Department and Institute for Strings, Cosmology, and Astroparticle Physics,\\
Columbia University, New York, NY 10027, USA
}%

\date{\today}

\begin{abstract}
The equivalence of inertial and gravitational masses is a defining feature 
of general relativity.
Here, we clarify the status of the equivalence principle for interactions 
mediated by a universally coupled scalar,
motivated partly by recent attempts to modify gravity at cosmological 
distances. Although a universal scalar-matter coupling is not mandatory,
once postulated, it is stable against classical and quantum renormalizations 
in the matter sector.
The coupling strength itself is subject to renormalization of course.
The scalar equivalence principle is violated only for objects for
which either the graviton self-interaction or the scalar 
self-interaction is important---the first applies to black holes,
while the second type of violation 
is avoided if the scalar is Galilean-symmetric.
\end{abstract}

\maketitle


The defining feature of general relativity (GR) 
is undoubtedly the equivalence principle, 
namely the equivalence of inertial and gravitational masses. 
Phrased in a field theoretic language pioneered by Weinberg
\cite{weinberg,weinbergGR}, general relativity
can be thought of as the unique (low energy) Lorentz-invariant 
theory of a massless spin-two field---the graviton.
Consider the motion of an object, be it big as a galaxy
or small as a proton, in a long-wavelength gravitational field---long 
compared to the object size.
How the object responds to the gravitational field is captured
by a coupling between the low energy graviton field and the
object's total physical energy-momentum. 
The equality of inertial and gravitational masses follows from
their deriving from the same physical energy-momentum.
We emphasize {\it physical}, meaning {\it renormalized}: the observed
mass of an object typically has many contributions on top of its 
constituents' rest masses, ranging from kinetic energies, to classical 
electromagnetic and gravitational potential energies, and 
to more genuinely quantum-mechanical corrections---
which are in fact the leading source of mass 
for nucleons, for instance.
Weinberg's argument guarantees that each renormalization---classical or 
quantum-mechanical, gravitational or 
otherwise---of an object's inertial mass is always accompanied by an 
{\em identical} renormalization of its gravitational mass. 
In other words, the universal coupling between the low energy graviton
and objects is robust against renormalizations.

This renormalization property of gravitational interactions is what makes the equivalence principle of GR robust, or more precisely, stable under renormalization group: if a system's constituents obey it individually, so does the system as a whole. 
Here
we wish to establish the analogous statement for universal 
long-range forces 
mediated by scalar fields---that is, how robust is a universally coupled 
scalar?
This question is far from academic, for most proposed theories that attempt to 
modify general relativity at cosmological distances, in one way or another 
involve a scalar degree of freedom that couples to the trace of the 
matter stress-energy tensor. This includes the 5-dimensional DGP model
\cite{DGP,LPR} and models of a massive/resonance graviton
\cite{AGS,gia,degrav}, not to mention classic scalar-tensor theories
\cite{Damour1992,Carroll:2003wy}.
We will see that the universal scalar coupling---a scalar 
equivalence principle---is robust against
renormalizations in the matter sector. We will also see that
violations arise only through nonlinearities in the graviton---the well known 
Nordtvedt effect \cite{Nordtvedt1968}---and nonlinearities in the scalar
which are crucial in recent theories for screening 
the scalar force on solar system scales \cite{us}.

Consider therefore a theory containing the gravitational field
$g_{\mu\nu}$, 
matter, and a scalar $\phi$ that couples to all forms of matter energy-momentum with the same strength:
\bea 
S  & = & 
S_{\rm EH} [g_{\mu\nu}] + S_{\rm m} [g_{\mu\nu}, \psi]  \nonumber \\
&& + \int \! d^4 x \sqrt{-g}  \big\{ {\cal L}_\phi [g_{\mu\nu}, \phi] + \phi \, T_{\rm m} [g_{\mu\nu}, \psi]  \big\} \; . \label{action}
\eea
$S_{\rm EH}$ is the Einstein-Hilbert action; $S_{\rm m}$ and $T^{\mu\nu}_{\rm m}$ are the action and stress-energy tensor for the matter degrees of freedom, collectively denoted by $\psi$.
 ${\cal L}_\phi$ encodes the dynamics of $\phi$; we are allowing for generic self-interactions. The coupling constant between $\phi$ and $T_{\rm m}$ is absorbed into the definition of $\phi$ itself (we will later discuss putting back an explicit
coupling constant when addressing the renormalization of its {\it value}.)
First of all, in comparison to GR, we already face a naturalness issue. 
The graviton's coupling to matter is uniquely dictated by symmetry \cite{weinberg, weinbergGR}. For $\phi$, there is no
symmetry enforcing this particular form of the coupling---$\phi$ may couple to any local scalar quantity built out of $\psi$. We can however postulate the $\phi T_{\rm m}$ form of the coupling, and check whether it is stable under classical and quantum-mechanical renormalizations. That is, the best we can hope for is {\it technical} 
naturalness.

We can then  ask, if $\phi T_{\rm m}$ is the correct coupling at some microscopic level, what is the coupling between a macroscopic or composite object and a long-wavelength 
$\phi$ field? 
The object is held together by internal forces, which may have any nature. 
For the moment, let us assume the
gravitational contributions to the object's total mass are negligible, 
though the object could still be gravitationally bound. Likewise, let's provisionally assume that $\phi$'s contribution to the total mass and $\phi$'s self interactions are also negligible.
We can thus set $g_{\mu\nu} \simeq \eta_{\mu\nu}$, and in the point-particle limit for the object (that is, at lowest order in the multipole expansion), the
scalar-object coupling is
\be \label{point_particle}
\int \! d \tau \, Q \, \phi \big( x^\mu(\tau) \big) \; ,
\ee
where $Q$ is the object's scalar charge, which in the object's rest frame reads
\be \label{Q}
Q \equiv \int d^3 x \, T_{\rm m} =  \int d^3 x \, \big( T^{00}_{\rm m} - T_{\rm m}^{ii} \big) \; .
\ee
The integral of $T^{00}_{\rm m}$ is just the total mass of the object, with no reference to how it is split into rest masses for the constituents, kinetic energies, and potential ones. The second term does not look  as nice.

However, we can rewrite $T^{ij}_{\rm m}$ as
\be
T_{\rm m} ^{ij}= \di_k \big( x^i T_{\rm m}^{kj} \big) -  x^i \,  \di_k  T_{\rm m}^{kj}  \; .
\ee
If we now integrate $T_{\rm m} ^{ij}$ over space, the first piece yields zero for a localized system. The second piece, using stress-energy conservation, can be rewritten as a {\em time}-derivative:
\be
\int d^3 x \, T_{\rm m} ^{ij} =  \di_t \int d^3 x \, x^i \, T_{\rm m}^{0j} 
\ee
(In fact by analogous arguments one can show that the integral of $T_{\rm m} ^{ij}$ is a {\em second} time-derivative---this is one incarnation of the tensor virial theorem.) 
Therefore, for stationary systems the spatial integral of $T_{\rm m} ^{ij}$ vanishes
\footnote{The same statement was derived in the context of solitons by Manton \cite{manton}.},
and for virialized systems it averages to zero on time-scales larger than the system's dynamical time. Equivalently, for a low-frequency external $\phi$ 
field, this $T_{\rm m} ^{ii}$
contribution to $Q$ is negligible with respect to the $T_{\rm m} ^{00}$ one.
We  are thus left with
\be
Q \simeq M \; ,
\ee
the total inertial mass of the object \footnote{If the object were a massless particle such as a photon, $Q=M=0$, so equivalence
principle is also obeyed.}. 

The above derivation shows that the equality of scalar charge and inertial mass
is robust against classical renormalization. 
The same proof applies essentially unaltered to quantum mechanical contributions to $Q$, at the {\em non-perturbative} level. Consider for instance a proton, $|p \rangle$. If at the microscopic level $\phi$ couples as in eq.~(\ref{action}) to quarks and gluons, the coupling between our proton and 
a long wavelength $\phi$ will be eq.~(\ref{point_particle}), with scalar charge
\be
Q = \langle p | \,  \textstyle{\int} d^3 x \, T_{\rm QCD} \,   | p \rangle \; ,
\ee
where $T^{\mu\nu}_{\rm QCD}(x) $ is the microscopic QCD stress-energy tensor operator, expressed in terms of quark and gluon fields. $T^{\mu\nu}_{\rm QCD}(x)$  is conserved as an operator. We can thus run the same algebra as in the classical case, and end up with
\be
Q = \langle p | \,  \textstyle{\int} d^3 x \, T^{00}_{\rm QCD} \,  | p \rangle 
- \di_t \langle p | \,  \textstyle{\int} d^3 x \, x^i \, T^{0i}_{\rm QCD} \,  | p \rangle \; .
\ee
The first term is the physical mass of the proton---by definition. The second term vanishes, because the proton is a non-perturbative stationary state of the QCD Hamiltonian.
We thus have
\be
Q = M  \; ,
\ee
like in the classical case.

Given its dryness, our proof deserves some comments. The crucial ingredient we are relying on is stress-energy conservation for matter  alone. 
As is clear from our action (\ref{action}), we are calling `matter' everything but the gravitational field and $\phi$ itself. 
We have thus demonstrated the robustness of the scalar equivalence principle against
renormalizations in the matter sector only.
What is exactly conserved is, of course, the total stress-energy (pseudo-)tensor 
$t_{\mu\nu}$ for matter, gravity, and $\phi$. 
This means that our result does not apply to systems where 
gravity or $\phi$ gives sizable contributions to the total mass, like a black hole.
Indeed, because of the no-hair theorem a black hole cannot couple to a long-wavelength scalar field, thus violating $Q \simeq M$---the 
Nordtvedt effect \cite{Nordtvedt1968,Will2006}. Note however that our result does apply to gravitationally or `scalar-ly' bound systems with negligible gravitational and scalar self-energy---in such a case $T_{\rm m} \simeq {t^\mu}_{\mu}$, and in our proof we could have just used $t_{\mu\nu}$, which is exactly conserved. More importantly,  our proof neglects contributions
to the scalar charge from $\phi$'s self-interactions, which as we mentioned are crucial for screening the scalar force in the solar system. Such self-interactions effectively renormalize the monopole coupling (\ref{Q}) of a localized  object to a long-wavelength $\phi$ by an amount
\be
\Delta Q = \int \! d^3 x \, \frac{\partial {\cal L}_\phi}{\partial \phi}[\phi_{\rm obj}] \; ,
\ee
where $\phi_{\rm obj}$ is the $\phi$ field dressing the object. Notice that here the derivative w.r.t.~$\phi$ is an ordinary one---not a functional one \cite{us}. This is because we want to isolate the monopole coupling, which involves the zero-momentum limit for $\phi$.
For generic  ${\cal L}_\phi$, this contribution can  yield violations of the equivalence principle of order one, such as 
in chameleon \cite{chameleon2} or symmetron \cite{symmetron} screening.
However there exists a class of observationally viable scalar self-couplings---those associated with Galilean invariance \cite{NRT}, used in Vainshtein screening \cite{Vainshtein}---that do not renormalize the scalar charge \cite{us}. For these interactions the equivalence principle is preserved as long as the gravitational and scalar binding energies are negligible.

\begin{figure*}[t]
\begin{center}
\includegraphics[width=4cm]{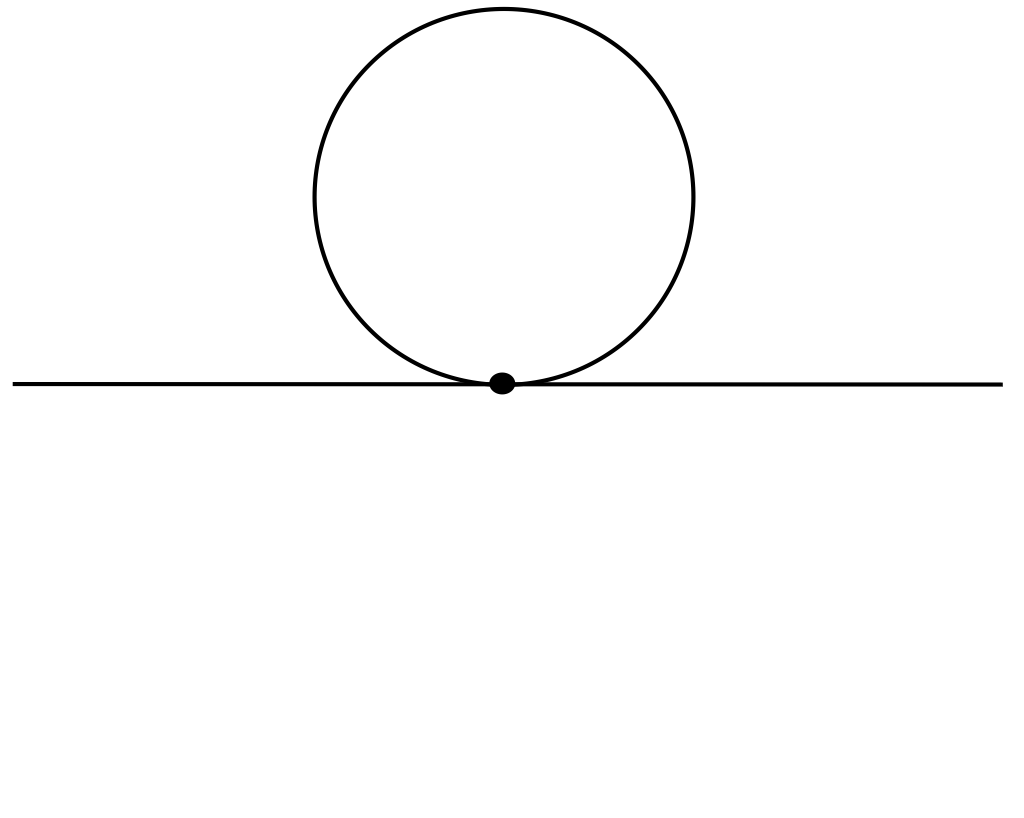} \hspace{.5cm}
\includegraphics[width=4cm]{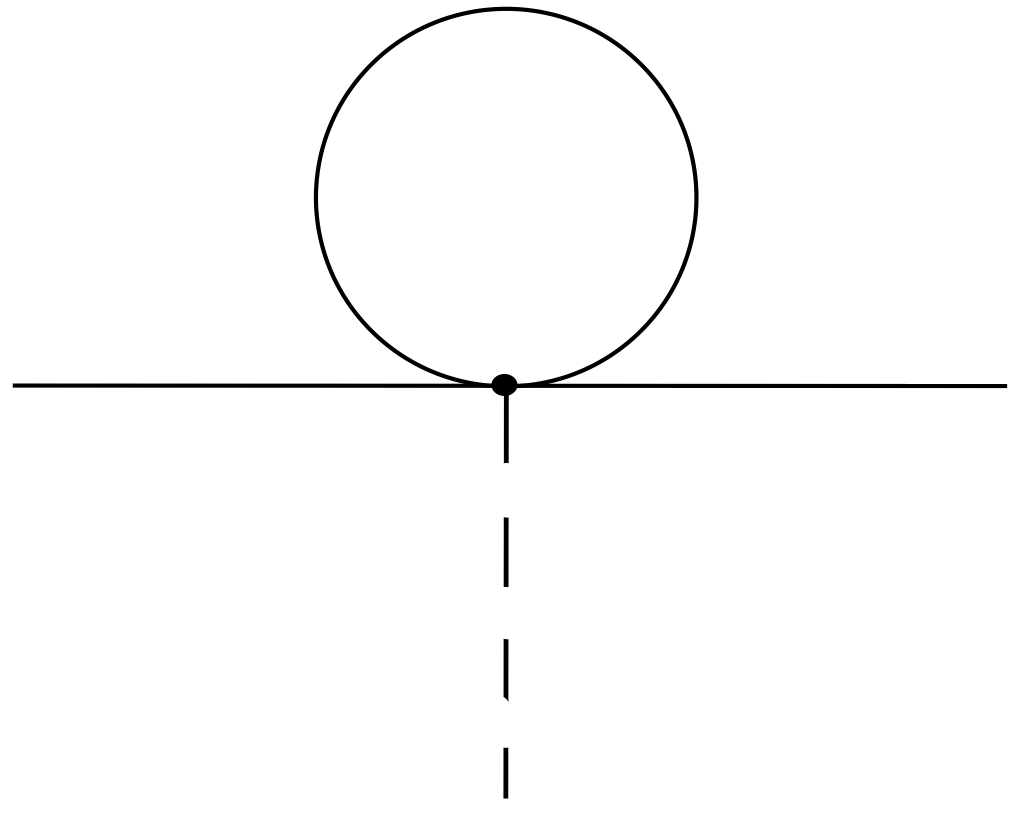}\hspace{.5cm}
\includegraphics[width=4cm]{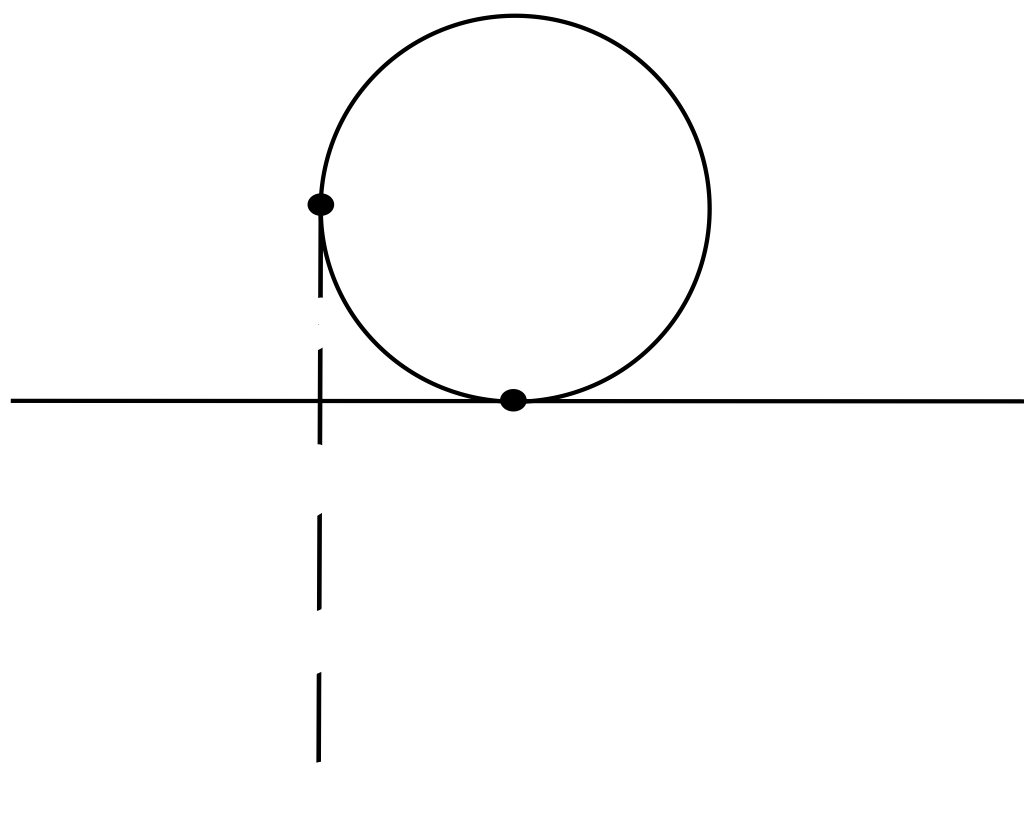}
\caption{\label{loop} \it One-loop contributions to the quadratic effective action for $\psi$ (solid lines) in the presence of an external $\phi$ (dashed line).}
\end{center}
\end{figure*}

Our simple derivation sheds light on what might otherwise appear to be miraculous cancellations in computations of quantum corrections to the scalar coupling. The robustness of the
universal coupling has been demonstrated by Fujii \cite{Fujii} in the context of a scalar coupled to
QED. Let's consider as another example
a matter sector with a set of interacting scalars $\psi_a$ with 
Lagrangian
\be \label{phi4}
{\cal L}_{\rm m} = \sum_a \sfrac12 \big[(\di \psi_a)^2 -  m_a^2 \, \psi_a^2\big] - \sum_{a,b}\lambda_{ab} \psi_a^2  \psi_b^2 \; .
\ee
For simplicity we are postulating a symmetry under $\psi_a \to - \psi_a$ for each particle species, so that quantum corrections do not generate kinetic and mass-mixings between different species. 
Let's couple our $\phi$ to the $\psi$s' stress-energy tensor, like in eq.~(\ref{action}).
We have
\be
T_{\rm m} = \sum_a \sfrac12 \big[-(d-2) (\di \psi_a)^2 + d\, m_a^2 \, \psi_a^2\big] + d \, \sum_{a,b}\lambda_{ab} \psi_a^2  \psi_b^2 \; ,
\ee
where $d$ is the spacetime dimensionality---we will use dimensional regularization for the UV divergences.
From fig.~\ref{loop}, we can compute the 1-loop contribution to the quadratic effective action for the $\psi$'s, in the presence of a long-wavelength external $\phi$. In fact the computation is made simpler  by 
noting that combining the matter Lagrangian with the interaction with $\phi$  we get the same structure as in eq.~(\ref{phi4}), but with $\phi$-dependent coefficients in front of each term. For a very long wavelength $\phi$, we can treat such coefficients as constant. Upon redefining $\psi_a \to \big[1+ \frac12 (d-2) \phi\big]\psi'_a $ to re-gain canonical normalization, we get
\bea
{\cal L}_{\rm m}  & + & \phi T_{\rm m}  \to \label{L_plus_phiT}\\
&& \sum_a \sfrac12 \big[(\di \psi'_a)^2 -  \hat m_a^2(\phi) \, \psi'_a{}^2\big] - \sum_{a,b}\hat \lambda_{ab} (\phi) \psi'_a{}^2  \psi'_b{}^2  \nonumber \; ,
\eea
where
\be
\hat m_a^2(\phi)  \equiv m_a^2 [1- 2 \phi] \; , \quad \hat \lambda _{ab}(\phi) \equiv \lambda_{ab}
[1+(d-4) \phi] \; ,
\ee
and we kept the linear order in $\phi$ only. It is now simple to retrieve quantum corrections to how $\phi$ couples to the $\psi$'s
---in the zero momentum limit for $\phi$---
from loop diagrams with no external $\phi$-lines. For instance, now the leftmost diagram of fig.~\ref{loop} is enough to produce the 1-loop effective action at quadratic order in the $\psi'$'s:
\be \label{deltaL}
\Delta {\cal L} = - \sfrac12 \,  \sum_{a}  \frac{\Delta m_a^2}{m_a^2} \, \hat m_a^2(\phi) \,  \psi'_a{}^2  \; ,
\ee
where
\be \label{deltam2}
\Delta m_a^2 \propto \sum_{b} \big( 2 \, \delta_{a\neq b} +12 \, \delta_{ab}\big)\, 
\lambda_{ab} \, \big[ \sfrac{1}{d-4} + \log (m_b/\mu) \big] \,  m_b^2  \, .
\ee
That is, for each particle species, the coupling to $\phi$ gets renormalized precisely by the same multiplicative factor as  the inertial mass, as predicted. Equivalently, undoing the field redefinition and expressing everything in terms of the original $\psi$, at quadratic order in $\psi$ the coupling with $\phi$ is still of the $\phi T_{\rm m}$ form:
\be \label{quantum_corrected_L}
{\cal L}_{\rm m} + \phi T_{\rm m} \to {\cal L}_{\rm} + \Delta {\cal L}_{\rm m } + \phi \,  (T_{\rm m} + \Delta T_{\rm m} ) \; ,
\ee
where $\Delta T_{\rm m}^{\mu\nu}$ is the correction to $T_{\rm m}^{\mu\nu}$ associated with $\Delta {\cal L}_{\rm m } $, i.e.
\be
\Delta T^{\mu\nu}_{\rm m} = \frac{2}{\sqrt{-g}} \frac{\delta \,  \Delta {\cal L}_{\rm m }} {\delta g_{\mu\nu}}\; .
\ee
Notice that this relies crucially on the mass-term quantum correction (\ref{deltaL})'s having the same universal $\phi$-dependence as the  tree-level mass term in (\ref{L_plus_phiT})---proportional to $(1 - 2\phi)$.
If the $\phi$-dependence were different from the tree-level one, and species-dependent, it would lead to two different kinds of particles falling at different rates.
The `miraculous cancellations' alluded to above, are here embodied by the equality
\be
\hat \lambda_{ab} (\phi) \, \big[ \sfrac{1}{d-4} + \log (\hat m_b(\phi)/\mu) \big] = \lambda_{ab} \, \big[ \sfrac{1}{d-4} + \log (m_b/\mu) \big] \; ,
\ee
which is valid at first order in $\phi$ and which was crucial to obtain the universal structure (\ref{deltaL}).
Given our general arguments above, we expect eq.~(\ref{quantum_corrected_L}) to hold at $\psi^4$ order as well.

This example thus illustrates and confirms our general result
for a universal scalar-matter coupling. The universality can be violated only when gravitational
or scalar self-interactions are important,
with the latter possibility
precluded in theories with Galilean symmetry.
In this regard, scalar forces are capable of obeying
an equivalence principle, though one not as strong and inevitable as that for the graviton.

We close with a few final remarks. First, our general analysis and our one-loop example show that starting with a $\phi T_{\rm m}$ coupling, upon renormalization the coupling between $\phi$ and matter will remain precisely $\phi T_{\rm m}$, if $T_{\rm m}$ is expressed in terms of the physical, renormalized masses and couplings. The overall coefficient in front does not get renormalized. However the Lagrangian describing the dynamics of $\phi$ does get renormalized---for instance matter loops with two $\phi$-external legs yield wave-function renormalizations for $\phi$. So, in the end, the universal coupling between matter and $\phi$-quanta---which we get by going to canonical normalization for $\phi$---does receive (universal) quantum corrections. 
More explicitly, a canonically normalized $\phi$ couples to matter as $\alpha \phi T_{\rm m}$, and the value of $\alpha$ is subject to renormalization. For instance, the value $\alpha=1/\sqrt{6}$ that defines $f(R)$ is not protected.
Perhaps more importantly, matter loops with an external $\phi$ and an external graviton will generically generate a kinetic mixing between the two fields. Demixing them from each other will not affect the universality of the scalar coupling to matter though, for the graviton is itself universally coupled. 

Second, our proof generalizes straightforwardly to a symmetric tensor field coupled to the matter stress-energy tensor, ${\cal L}\supset X_{\mu\nu} \, T_{\rm m}^{\mu\nu}$, of which our scalar coupling is just a special case with $X_{\mu\nu} = \phi \,\eta_{\mu\nu}$. Indeed the point-particle coupling (\ref{point_particle}) generalizes to
\be \label{point_particle_tensor}
\int \! d \tau \, Q^{\mu\nu} \, X_{\mu\nu} \big( x^\mu(\tau) \big) \; ,
\ee
where the `tensor charge' $Q^{\mu\nu}$ is defined as
\be \label{Q_tensor}
Q^{\mu\nu} \equiv \int d^3 x \, T^{\mu\nu}_{\rm m}  \; ,
\ee
in the object's rest frame. $T_{\rm m}^{ij}$ integrates to zero because of the same reasons (and under the same assumptions) as above; $T_{\rm m}^{0i}$ integrates to the total momentum, which also vanishes in the object's rest frame. We thus have that in the rest frame the only non-vanishing entry of $Q^{\mu\nu}$ is $Q^{00} = M$---the total inertial mass of the object. Going to a generic frame we thus get that the coupling (\ref{point_particle_tensor}) reduces to
\be
M \int \! d \tau \, u^\mu u^\nu \, X_{\mu\nu} \big( x^\mu(\tau) \big)  \; ,
\ee
where $u^\mu$ is the objects four-velocity. This is precisely how the gravitational field couples to an object in the point-particle approximation.
Clearly the same proof applies in the quantum-mechanical case, with the same modifications as above.

Finally, in the quantum-mechanical case we have been neglecting loop contributions with $\phi$ and graviton internal lines. These are suppressed by inverse powers of the Planck mass (assuming the scalar couples to matter with gravitational strength), and can be safely neglected as long as
matter self-interactions are much stronger than gravity.

\noindent
{\em Acknowledgements.}
We would like to thank Niayesh Afshordy, Cristian Armend\'ariz-Pic\'on, and especially Eduardo Ponton for useful discussions. This work is supported in part by the DOE (DE-FG02-92-ER40699) and NASA ATP
(09-ATP09-0049).



\begin{thebibliography}{99}

\bibitem{weinberg}
  S.~Weinberg,
  Phys.\ Rev.\  {\bf 135}, B1049 (1964).

\bibitem{weinbergGR}
  S.~Weinberg,
  Phys.\ Rev.\  {\bf 138}, B988 (1965).

\bibitem{DGP}
  G.~R.~Dvali, G.~Gabadadze and M.~Porrati,
  Phys.\ Lett.\  B {\bf 485}, 208 (2000)
  [arXiv:hep-th/0005016].

\bibitem{LPR}
  M.~A.~Luty, M.~Porrati and R.~Rattazzi,
  JHEP {\bf 0309}, 029 (2003)
  [arXiv:hep-th/0303116].

\bibitem{AGS}
  N.~Arkani-Hamed, H.~Georgi and M.~D.~Schwartz,
  Annals Phys.\  {\bf 305}, 96 (2003)
  [arXiv:hep-th/0210184].

\bibitem{gia}
  G.~Dvali,
  Modified Gravity,''
  New J.\ Phys.\  {\bf 8}, 326 (2006)
  [arXiv:hep-th/0610013].

\bibitem{degrav}
  G.~Dvali, S.~Hofmann and J.~Khoury,
  Phys.\ Rev.\  D {\bf 76}, 084006 (2007)
  [arXiv:hep-th/0703027].

\bibitem{Damour1992}
  T.~Damour and G.~Esposito-Farese,
  Class.\ Quant.\ Grav.\  {\bf 9}, 2093 (1992).

\bibitem{Carroll:2003wy}
  S.~M.~Carroll, V.~Duvvuri, M.~Trodden and M.~S.~Turner,
  Phys.\ Rev.\  D {\bf 70}, 043528 (2004)
  [arXiv:astro-ph/0306438].


\bibitem{Nordtvedt1968}
K. Nordtvedt,
  Phys. Rev. {\bf 169}, 1014 (1968).

\bibitem{us}
  L.~Hui, A.~Nicolis and C.~Stubbs,
  Phys.\ Rev.\  D {\bf 80} (2009) 104002
  [arXiv:0905.2966 [astro-ph.CO]].


\bibitem{manton}
N.~S.~Manton
  J.\ Math.\ Phys.\  {\bf 50}, 032901 (2009)
  [arXiv:0809.2891 [hep-th]].

\bibitem{Will2006}
C. M. Will,
  Living Reviews in Relativity {\bf 9}, 3 (2006)
 [arXiv:gr-qc/0510072].

\bibitem{chameleon2}
  J.~Khoury and A.~Weltman,
  Phys.\ Rev.\ Lett.\  {\bf 93}, 171104 (2004)
  [arXiv:astro-ph/0309300].

\bibitem{symmetron}
  K.~Hinterbichler and J.~Khoury,
  Restoration,''
  arXiv:1001.4525 [hep-th].

\bibitem{NRT}
  A.~Nicolis, R.~Rattazzi and E.~Trincherini,
  Phys.\ Rev.\  D {\bf 79}, 064036 (2009)
  [arXiv:0811.2197 [hep-th]].

\bibitem{Vainshtein}
  A.~I.~Vainshtein,
  Phys.\ Lett.\  B {\bf 39}, 393 (1972).

\bibitem{Fujii}
  Y.~Fujii,
  Mod.\ Phys.\ Lett.\  A {\bf 12}, 371 (1997)
  [arXiv:gr-qc/9610006].





\end{thebibliography}
\end{document}